\documentclass[aps,prl,twocolumn,superscriptaddress,showpacs]{revtex4}

\usepackage{amsmath}
\usepackage{graphicx}
\usepackage{dcolumn}
\usepackage{sidecap}
\newcommand{\mypath}[1]{./#1}

\begin{document}

\title{Competing Ferromagnetic and Charge-Ordered States
in Models for Manganites: \\
the Origin of the CMR Effect}

\author{Cengiz \c{S}en}
\affiliation{National High Magnetic Field Laboratory and Department of Physics,
Florida State University, Tallahassee, FL 32310}

\author{Gonzalo Alvarez}
\affiliation{Computer Science \& Mathematics Division
and Center for Nanophase Materials Sciences, 
Oak Ridge National Laboratory, Oak Ridge, TN 37831}

\author{Elbio Dagotto}
\affiliation{Materials Science and Technology Division, Oak Ridge National Laboratory, Oak Ridge, TN 32831}
\affiliation{Department of Physics and Astronomy, The University of Tennessee, Knoxville, TN 37996}

%\affiliation{%

%\author{...}
%\author{...}
%\author{...}
%\affiliation{National High Magnetic Field Lab and Department of Physics,
%Florida State University, Tallahassee, FL 32310}

\date{\today}

\begin{abstract}
The one-orbital model for manganites with cooperative phonons and superexchange coupling $J_{\rm AF}$
has been investigated via large-scale Monte Carlo (MC) simulations. Results for two-orbitals are also 
briefly discussed. Focusing on electronic density $n$=0.75, a regime of competition between ferromagnetic (FM)
metallic and charge-ordered (CO) insulating states was identified. In the vicinity of the associated 
bicritical point, colossal magnetoresistance (CMR) effects were observed. 
%The CMR magnitude is much 
%larger than recently reported when randomly distributed polarons form the competing insulator. 
%The 
%appearance of 
The CMR is associated with the development of short-distance correlations 
among polarons, above the spin ordering temperatures, resembling the charge arrangement of the 
low-temperature CO state.
\end{abstract}


\maketitle

{\it Introduction:}
The colossal magnetoresistance of the manganites 
is an example of the complex behavior and nonlinearities that can
emerge in materials where several degrees of 
freedom are simultaneously active~\cite{otherreviews,review}.
Understanding theoretically the CMR effect is important not only for its intrinsic
value in the Mn oxide context, but also
to provide a paradigm for rationalizing a plethora of related
complex phenomena unveiled
in several other transition metal oxides in recent years~\cite{science}.
The current scenarios to understand the CMR behavior rely on
the existence of competing states and the emergence of nanometer-scale electronic
structures~\cite{review}. Using simple models of phase competition (valid
at large length scales) and resistor-network approximations, 
CMR behavior was observed~\cite{burgy}.
However, 
for a deeper understanding, it is imperative to unveil the CMR
effect in more fundamental microscopic models, 
using fully unbiased many-body techniques to handle the strong interactions.
Along this line, considerable progress 
was recently made with the report of large magnetoresistance
effects in Mn-oxides models in the case
where the double-exchange (DE)
induced FM metal competes with a FM insulator made out of 
randomly localized polarons~\cite{verges,kumar,sen06}.
However, a complete rationalization of CMR physics needs a CO
antiferromagnetic (AF) state as the true 
competitor of the FM metal, since it  is under these
circumstances that the magnetoresistance effect is truly colossal~\cite{otherreviews,review}. For this goal, it is crucial to incorporate
the $t_{\rm 2g}$ AF superexchange $J_{\rm AF}$, 
known to be fundamental to reproduce the CO-AF states observed 
in real manganite phase diagrams~\cite{review}.

The main purpose of this manuscript is to present 
large-scale computational studies
of models for manganites, incorporating now explicitly the $J_{\rm AF}$ interaction.
The main result is that a CMR effect, with CMR
ratios as large as 10,000\%, is unveiled
using realistic models and electronic densities, and unbiased methods. 
The $J_{\rm AF}$ coupling is shown to be crucial for the
magnitude of the effect.
A simple picture for the origin of the CMR effect is discussed,
that relies on nanometer-scale short-range 
order above the Curie temperature $T_{\rm C}$, in qualitative
agreement with previously proposed mixed-phase scenarios.
The comprehensive investigation reported here was possible 
by using hundreds of nodes of the Cray XT3 supercomputer operated by the National Center for Computational Studies, at Oak
Ridge National Laboratory (ORNL). 
%, exploiting the supercomputer
%low latency and scalability. 
The same effort on conventional PC clusters would have demanded
several years for completion.

{\it Models and Techniques:} 
The models and methodology used in this effort
have been extensively discussed before~\cite{sen06,review}.
Then, only a brief description is
here included. The one-orbital Hamiltonian is given by $H$=$H_{\rm 1b}$$+$$H_{\rm AF}$,
where $H_{\rm 1b}$ contains the standard fermionic 
nearest-neighbor (NN) hopping term at infinite Hund's coupling, plus the
interaction with the oxygen phonons which are assumed classical. 
The full $H_{\rm 1b}$ term, including on-site disorder, 
is explicitly defined in Eq.(2) of Ref.~\onlinecite{sen06}. The two-orbital model, also used here,
is identical to Eq.(4) of Ref.~\onlinecite{sen06}. 
The electron-phonon
coupling is denoted by $\lambda$, the strength of the on-site quenched disorder is $\Delta$, the chemical
potential $\mu$ regulates the density $n$, and the hopping $t$=1
is the energy unit~\cite{hotta-coul}. In this study, the second term
$H_{\rm AF}$=$J_{\rm AF}\sum_{\langle ij \rangle} \vec{S}_i \cdot \vec{S}_j$ plays a key role.
This interaction represents  the NN superexchange, with coupling $J_{\rm AF}$,
between the (assumed classical) $t_{\rm 2g}$ spins $\vec{S}_i$.
The study of this model involves the 
standard exact diagonalization of the fermionic sector for a given
classical phononic and $t_{\rm 2g}$ spins configuration, 
updated via a MC procedure~\cite{review}.
The charge-charge correlations were evaluated using $C(j)$=$(1/N)\sum_{i}(\langle n_{i}n_{i+j}
\rangle-\langle n \rangle^{2})$ with $n_{i}$ the electronic density at site $i$, 
and $N$ the total number of sites.
The resistivity $\rho$ has been calculated by taking the inverse of the mean conductivity $\sigma$, 
where the latter is related to the conductance $G$ by $G$=$\sigma L^{d-2}$, $d$ being the dimension, and
$L$ the linear size of the lattice. The conductance $G$ was obtained
using the Kubo formula~\cite{verges00}. For further details, the reader should consult Sec.~II 
of Ref.~\onlinecite{sen06}. Finally, note that finite clusters do not 
lead to true singularities, such as a critical
temperatures. However,
decades of investigations in a wide variety of contexts 
have shown that finite systems can accurately capture
the rapid growth of correlation lengths in narrow temperature ranges 
expected from criticality.
These temperatures can be safely associated with
true critical temperatures, 
and this is the convention followed here and in many
previous studies~\cite{review,burgy,verges,kumar,sen06}.

\begin{figure}
\centerline{
\includegraphics[clip,width=7.5cm, height=7.5cm]{\mypath{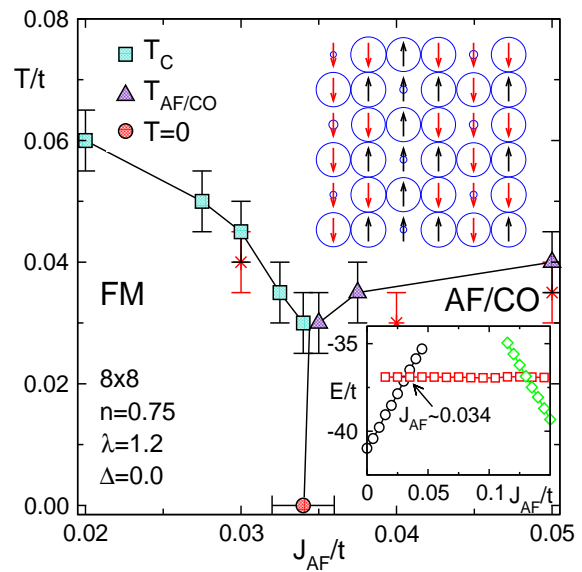}}}
\caption{(Color online) Clean limit 
MC phase diagram using $8$$\times$$8$ and 12$\times$12 lattices, 
at $n$=$0.75$ and  $\lambda$=$1.2$. 
The AF/CO state is schematically shown, with the radius of the circles proportional to the electronic density, and arrows
representing the $t_{\rm 2g}$ spins. Charge is uniform in the competing FM state. 
At each temperature, $1$$\times$$10^5$ thermalization and 
$5$$\times$$10^4$ measurement MC steps were carried out for the 8$\times$8 clusters (and $\sim$$7500$
and $5000$, respectively, for the 12$\times$12 cluster points indicated by red stars).
Random starting spin configurations were used in all simulations, but the results have also 
been checked for convergence using ordered starting configurations. $T_{\rm C}$ is defined as the
temperature where the spin-spin correlations at the maximum cluster 
distance nearly vanish. Similarly, $T_{\rm AF/CO}$ is
the temperature where both AF and FM correlations vanish at the maximum allowed distances. 
At $T_{\rm AF/CO}$, the charge structure factor $n(\mathbf{q})$ at $\mathbf{q}=\mathbf{q}_{\rm CO}$ also
vanishes. $Inset$: energy vs. $J_{\rm AF}$ at very low $T$$\sim$0, with the FM (CO/AF) phase in black (red).
Green dots indicate a G-ordered AF regime. 
%Thearrow in the inset indicates the low temperature critical $J_{\rm AF}$ for the FM-AF/CO transition. 
Here, the spins were frozen to their expected states and the phonons were MC relaxed. 
%Other intersection point at $J_{\rm AF}$$\sim$$0.13$ indicates the AF/CO-G(AF) transition.
}
\label{Figure1}
\end{figure}

{\it Clean-Limit Phase Diagram:} The key novelty of this investigation
is the existence of an insulating CO/AF state competing with the FM metal to generate the CMR. 
%as opposed  to an insulator with localized polarons~\cite{verges,kumar,sen06}. 
In Fig.~\ref{Figure1}, the clean-limit $\Delta$=0 
one $e_{\rm g}$-orbital model phase diagram is presented, at an experimentally realistic density
$n$=0.75 and intermediate $\lambda$=1.2. At $J_{\rm AF}$=0, the ground state is a DE
generated standard FM metal, with 
charge uniformly distributed. With increasing $J_{\rm AF}$,
a first-order transition occurs to a CO/AF state schematically 
shown in Fig.~\ref{Figure1}~\cite{bicritical}. 
The holes are arranged regularly, separated by distances $2$ and $\sqrt{5}$. This state is degenerate
with a state rotated in 90$^o$. The spins are also indicated.
This type of states are
the analog of the more realistic 
CO/AF states that exist in two-orbital Mn-oxide models, such as 
the CE state~\cite{review}.

\begin{figure}
\centerline{
\includegraphics[clip,width=7.5cm]{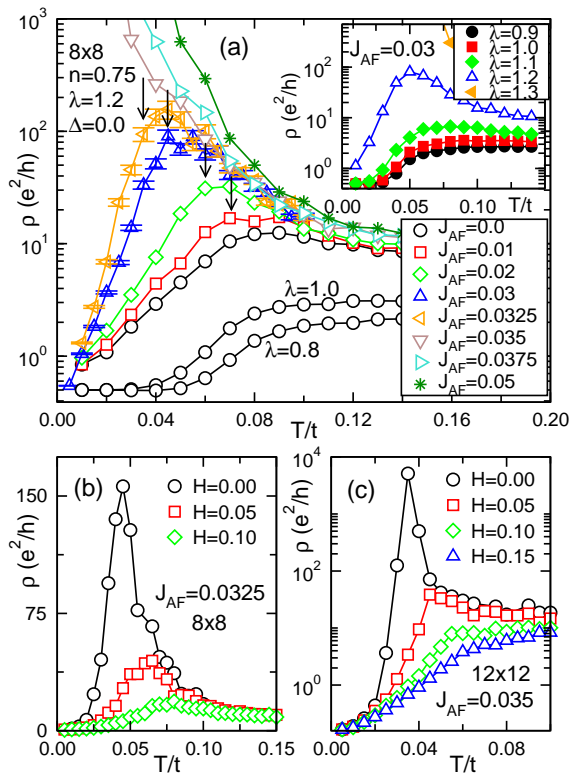}}
\caption{(Color online) Resistivity $\rho$ vs $T$ curves for various parameters: (a) Fixing $\lambda$=$1.2$ 
and varying $J_{\rm AF}$. Arrows indicate $T_{\rm C}$'s. Results at $\lambda$=$0.8$ and $\lambda$=$1.0$, with 
$J_{\rm AF}$=$0.0$, are also shown. $Inset$: results fixing $J_{\rm AF}$=$0.03$ and varying $\lambda$.
(b) Effect of magnetic fields (indicated, in $t$ units) on $\rho$ using $J_{\rm AF}$=$0.0325$, on an $8$$\times$$8$ lattice. 
(c) Same as (b) but for $J_{\rm AF}$=$0.035$, on a $12$$\times$$12$ lattice. In (a) and (b), 
MC steps and the starting configurations are the same as in Fig.~\ref{Figure1}. In (c), $7500$
thermalization and $5000$ measurement steps were used. Typical error bars are indicated in (a).}
\label{Figure2}
\end{figure}

{\it Clean-limit CMR:} The most important result of these investigations is shown in Fig.~\ref{Figure2},
where the $\rho$ vs. temperature $T$ curves are shown. Consider Fig.~\ref{Figure2}(a): here $\rho$ vs. $T$
presents the expected insulating behavior at large $J_{\rm AF}$, and the canonical bad-metal 
DE form at small $J_{\rm AF}$ (or reducing $\lambda$). The remarkable result appears in
between, mainly in the narrow $J_{\rm AF}$ interval approximately between 0.02 and 0.0325. In this regime,
$\rho$ vs. $T$ presents a canonical CMR shape, with insulating behavior at large $T$, transforming into
a broad high peak upon cooling (logarithm scale used), followed by metallic behavior at low $T$.
The $T_{\rm C}$ is approximately located at the resistivity peak.
This is in agreement with the experimental phenomenology of CMR manganites, and with
the theoretical scenarios~\cite{burgy,review} based on a CMR emerging from phase competition between
a FM metal and a CO-AF insulator. The result is particularly
remarkable considering the relatively small clusters used in our studies: the origin of the
CMR effect unveiled here must lie in phenomena occurring at the
nanometer scale, as recently remarked~\cite{mathieu}. 
The only ``price'' to pay is
the tuning of couplings: the CMR shape is obtained in narrow intervals $\Delta \lambda$$\sim$0.1
(inset Fig.~\ref{Figure2}(a)) and $\Delta J_{\rm AF}$$\sim$0.02. 
However, this delicate tuning is
removed by adding quenched disorder, 
as shown before~\cite{burgy,kumar,sen06}, and also
below in the present effort.

In Figs.~\ref{Figure2}(b,c), the effects of magnetic fields $H$ are shown. 
In agreement with experiments, the peak in $\rho$ is strongly suppressed by magnetic
fields that are small in the natural units $t$=1. The concomitant 
CMR ratios $[(\rho(0)-\rho(H))$/$\rho(H)]$$\times$100\% are as large as 10,000\%. The comparison
between 8$\times$8 and 12$\times$12 clusters 
also show that size effects are mild in these investigations.

\begin{figure}
\centerline{
\includegraphics[clip,width=8cm]{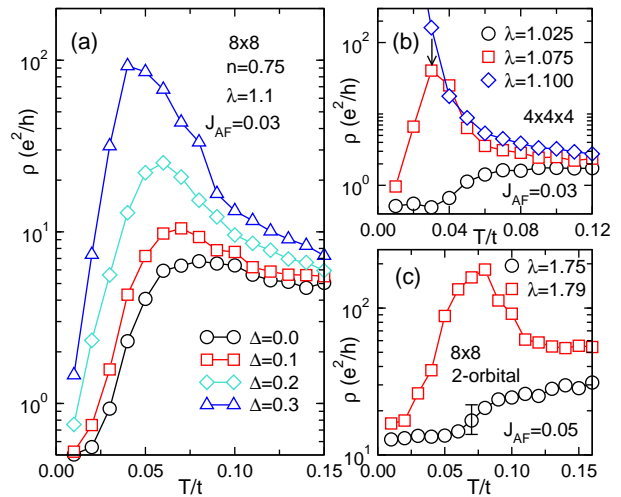}}
\caption{(Color online) (a) $\rho$ vs. $T$ in the presence of quenched disorder $\Delta$. 
Up to ten different disorder realizations were used in calculations with quenched disorder.
Only small changes between configurations were observed.
% and it is shown
%that the results are similar in all. Shown are the results corresponding to one disorder configuration.
MC steps and starting configurations are as in Fig.~\ref{Figure1}.
(b) $\rho$ vs. $T$ using a $4$$\times$$4$$\times$$4$ lattice, parametrized with $\lambda$,
at $J_{\rm AF}$=$0.03$. (c) Two orbitals $\rho$ vs. $T$ results using an $8$$\times$$8$ lattice
for $J_{\rm AF}$=$0.05$. In (b) and (c), $4000$ thermalization and $4000$ measurement MC steps were used,
$n$=0.75, and the clean limit $\Delta$=0 was studied.}
\label{Figure3}
\end{figure}

{\it Influence of Quenched Disorder, Dimensionality, and Number of Orbitals:} Fig.~\ref{Figure3}(a) shows typical 
$\rho$ vs. $T$ curves in the presence of on-site-energy quenched disorder. 
As discussed before~\cite{sen06,kumar}, the disorder enhances the tendency toward having
$\rho$ vs. $T$ curves with the canonical CMR shape, and
the fine tuning problem is avoided. This is important to rationalize the universality of the CMR effect,
present in so many Mn oxides that they cannot all be fined tuned to the same couplings.

To complete our investigations, the effects of dimensionality and orbital number were also addressed.
In Fig.~\ref{Figure3}(b) results are shown using a 4$\times$4$\times$4 cluster, as a representative of
a 3D lattice, in the clean limit. The $\rho$ vs. $T$ curves are very similar to those found in 2D systems.
%suggesting the absence of strong dimensionality effects. 
In Fig.~\ref{Figure3}(c), results using a two-orbital
model are presented, with similar conclusions. 
Figure~\ref{Figure3}(c) and others not shown,
suggest that the results are qualitatively the same for the one- and two-orbital models~\cite{2-orbitals}.

\begin{figure*}
\begin{minipage}{12cm}
\begin{center}
\includegraphics[width=11.5cm,clip]{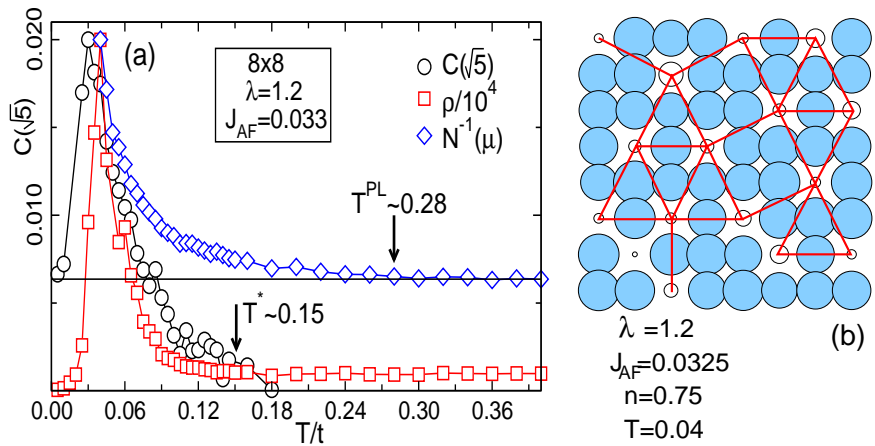}
\end{center}
\end{minipage}
\hfill
\begin{minipage}{5cm}
\begin{center}
\caption{(Color online) (a) 
%%Charge correlations at 
%%distance $\sqrt{5}$ (labeled C($\sqrt{5}$)) for the parameters $\lambda$ and $J_{\rm AF}$
%%indicated, and at temperature $T$=$0.06$$>$$T_{\rm C}$. 
%%(b) 
MC averaged C($\sqrt{5}$) vs. $T$, showing a qualitative
similarity with the rescaled resistivity (also shown). This agreement occurs below the $T^*$ indicated.
At higher $T$, $\rho$ is flat and C$(\sqrt{5})$ nearly vanishes. Also shown is the inverse
of $N(\omega=\mu)$, to indicate the PG formation at $T^{pl}$. 
%%Results corresponding to randomly chosen individual MC snapshot are also shown: they present
%%a similar, but enhanced, 
%%behavior as the averages. 
(b) Typical MC snapshot with the radius of the circles proportional to the local charge density. 
Also shown are the hole-hole 
distances $\sqrt{5}$ and $2$ of relevance (see text). 
%%In (a) and (b), the MC steps and 
%%starting configurations used are the same as in Fig.~\ref{Figure1}. 
}
\label{Figure4}
\end{center}
\end{minipage}
\end{figure*}

{\it Understanding the CMR Effect:}  
To understand the CMR effect found here,
it is important to analyze short-distance charge correlations above the ordering temperatures. 
%Equilibrated MC snapshots are also helpful to intuitively unveil
%the CMR physics.
In particular, the correlations at distance $\sqrt{5}$, denoted $C(\sqrt{5})$, are important since 
they are robust in wide ranges of temperatures and couplings, they 
are ubiquitous in most MC snapshots, and of course they are
also characteristic of the CO/AF state in Fig.~\ref{Figure1}.
%In Fig.~\ref{Figure4}(a), $C(\sqrt{5})$ is shown at high temperature,
%from a regime far from charge-ordering to couplings where the CO/AF state 
%becomes stable at low temperature. The
%smooth behavior of $C(\sqrt{5})$  reveals a slow build up of charge ordering in the normal state. 
%This is correlated with the increase of $\rho$ at fixed $T$ (Fig.~\ref{Figure2}). 
A clear relation between $C(\sqrt{5})$ and $\rho$ is
shown in Fig.~\ref{Figure4}(a), where they are plotted together for a representative $J_{\rm AF}$. 
In the regime below a new characteristic temperature
$T^*$$\sim$0.15, both quantities nicely track each other, 
showing that the increase of short-distance charge correlations causes the 
increase in $\rho$ upon cooling.
This figure is conceptually similar to Fig.~4 of Ref.~\onlinecite{lynn}, 
where $\rho$ for La$_{0.7}$Ca$_{0.3}$MnO$_3$ was 
shown together with the neutron scattering
$(1/4,1/4,0)$ polaron peak that describes 
short-distance polaron correlations. Such a ``correlated polaron'' signal is
the analogous of short-range CO in our one-band model, 
a precursor of the CO state stabilized at low $T$ close in parameter space.
Figure~\ref{Figure4}(a) shows that
the short-distance charge order is reduced at low $T$ 
but it does not disappear: in this regime 
the FM order of the spins, together with some reduction in charge correlations,
causes the dramatic resistance drop.

The scale $T^*$ where short-range correlations start to build up 
was already predicted based on
phenomenological investigations~\cite{burgy}, it was also found 
experimentally~\cite{argyriou}, but had not been 
reported before using basic models for
manganites. The resistivity becomes nearly flat above $T^*$.
In our investigations, a pseudogap (PG) in the density-of-states $N(\omega)$ was also observed~\cite{PG}, and its
inverse at $\mu$ is shown in Fig.~\ref{Figure4}(a). The PG disappears at a higher temperature $T^{pl}$, probably
related to polaron formation. The resistivity is not affected by $T^{pl}$, but upon further cooling
to $T^*$ those polarons become correlated and $\rho$ dramatically increases~\cite{billinge}.

%\begin{figure}
%\centerline{
%\includegraphics[clip,width=7cm]{\mypath{Figure4.eps}}}
%\caption{(Color online) (a) 
%%Charge correlations at 
%%distance $\sqrt{5}$ (labeled C($\sqrt{5}$)) for the parameters $\lambda$ and $J_{\rm AF}$
%%indicated, and at temperature $T$=$0.06$$>$$T_{\rm C}$. 
%%(b) 
%MC averaged C($\sqrt{5}$) vs. $T$, showing a qualitative
%similarity with the rescaled resistivity (also shown). This agreement occurs below the $T^*$ indicated.
%At higher $T$, $\rho$ is flat and C$(\sqrt{5})$ nearly vanishes. Also shown is the inverse
%of $N(\omega=\mu)$, to indicate the PG formation at $T^{pl}$. 
%%Results corresponding to randomly chosen individual MC snapshot are also shown: they present
%%a similar, but enhanced, 
%%behavior as the averages. 
%(b) Typical MC snapshot with the radius of the circles proportional to the local charge density. 
%Also shown are the hole-hole 
%distances $\sqrt{5}$ and $2$ of relevance (see text). 
%%In (a) and (b), the MC steps and 
%%starting configurations used are the same as in Fig.~\ref{Figure1}. 
%}
%\label{Figure4}
%\end{figure}

Finally, while averaged quantities are certainly crucial for quantitative studies, the ``by eye'' examination of
MC snapshots is also illustrative. In Fig.~\ref{Figure4}(b), a typical snapshot in the CMR regime is shown.
Pairs of holes located at the distances $\sqrt{5}$ and $2$ are highlighted. Clearly,
MC equilibrated configurations do $not$ have randomly distributed holes (i.e., it is not a gas
of heavy polarons), but special distances are preferred over others: the polarons are correlated.
Those distances are the same that characterize 
the low-$T$ CO state (Fig.~\ref{Figure1}). Also, in the MC snapshots the $t_{\rm 2g}$ spins at the hole locations 
and their
four neighbors were found to be polarized similarly to that shown
in Fig.~\ref{Figure1}.
Thus, puddles of the CO-AF state 
appear in the MC snapshots, and their existence is correlated with the
shape of the $\rho$ vs. $T$ curves.

{\it Conclusions:}  The results reported here show that realistic models for manganites -- with
the DE, electron-phonon, and $J_{\rm AF}$ interactions -- can explain 
the CMR effect and its magnitude. The coupling $J_{\rm AF}$ is crucial, since it is needed
to stabilize the CO-AF state that competes with the FM metal. The origin of the CMR is the formation
of nano-scale regions above 
$T_{\rm C}$,  with the same charge and spin pattern as the low-$T$ insulating CO-AF state.
To obtain CMR effects, clusters
of just a few lattice 
spacings in size appear sufficient~\cite{mathieu}. Our results show that the CMR effect is much
larger when the insulating competing state has charge ordering tendencies, 
as opposed to having merely randomly localized polarons.
The latter state has been recently identified using STM techniques, in a bilayered $x$=0.3 manganite 
with mild in-plane magnetoresistance~\cite{aeppli}.
If the same STM experiment is repeated at $x$=0.4, where the CMR is much stronger, it is predicted
that {\it aggregates} of polarons resembling a CO state should be observed above $T_{\rm C}$.

{\it Acknowledgment:}
We acknowledge discussions with D. Argyriou, S. Kumar, 
J. W. Lynn, P. Majumdar, and R. Mathieu,
and support via the NSF grant DMR-0443144 and the LDRD program at ORNL.
Most of the computational work was performed on the Cray XT3
of the Center for Computational Science at 
ORNL, managed by UT-Battelle, LLC, for the 
U.S. Department of Energy under Contract DE-AC05-000R22725.
%This work is also supported by the LDRD
%program at ORNL and by the NSF grant DMR-0443144.
This research used the SPF computer program and software toolkit developed at ORNL 
(http://mri-fre.ornl.gov/spf).

\end{document}